\documentclass[preprint2]{aastex}
\usepackage{graphicx}

\shorttitle{Pre-flare signatures}
\shortauthors{Kors\'os et al.}

\begin{document}

\title{Pre-flare dynamics of sunspot groups}

\author{M. B. Kors\'os, T. Baranyi and A. Ludm\'any}
\affil{Heliophysical Observatory, Research Centre for Astronomy and Earth Sciences, Hungarian Academy of Sciences,
\\4010 Debrecen, P.O. Box 30, Hungary}
\email{korsos.marianna@csfk.mta.hu}
\email{baranyi.tunde@csfk.mta.hu}
\email{ludmany.andras@csfk.mta.hu}

\begin{abstract}

Several papers provide evidences that the most probable sites of flare onset are the regions of high horizontal magnetic field gradients in solar active regions. Besides the localization of flare producing areas the present work intends to reveal the characteristic temporal variations in these regions prior to flares. This study uses sunspot data instead of magnetograms, it follows the behaviour of a suitable defined proxy measure representing the horizontal magnetic field gradient. The source of the data is the SDD (SOHO/MDI-Debrecen Data) sunspot catalogue. The most promising pre-flare signatures are the following properties of the gradient variation: i) steep increase, ii) high maximum, iii) significant fluctuation and iv) a gradual decrease between the maximum and the flare onset which can be related to the "pull mode" of the current layer. These properties may yield a tool for the assessment of flare probability and intensity within the next  8--10 hours.

\end{abstract}

\keywords{Sun:flares, sunspots}

\section{Introduction}

The source of the released energy in flares is the free magnetic energy, it is provided by the nonpotential or current carrying component of the active region magnetic field \citep{priest2002}. The large current layers cannot be observed directly, their localization needs the detailed knowledge of the magnetic field structure by extrapolating the data of surface magnetic fields to the corona, this is a complicated and time consuming task. Therefore the recent attempts focus on directly observable signatures of nonpotentiality to find diagnostically reliable pre-flare properties. 

The most promising observable feature is the inversion line separating the two areas of opposite polarities in the active regions. \citet{schrijver2007} demonstrated on a large sample that the occurrences of intensive flares are connected to inversion lines of high magnetic gradient. \citet{mason2010} defined a parameter named GWILL (gradient-weighted inversion line length) and found that it increased significantly prior to flares. They admitted, however, that GWILL is not suitable for real time prediction. \citet{cui2006} considered the maximum horizontal gradient and the length of the neutral line. Along with the neutral line and horizontal gradient the following additional features have been studied as possible pre-flare signatures: length of strong-shear main neutral line, net electric current arching from one polarity to the other, and a flux-normalized measure of the field twist \citep{falcon2002}; total magnetic energy dissipated in a unit layer per unit time \citep{jing2006}; shear \citep{cui2007}; effective connected magnetic field \citep{georg2007};  weighted (integrated) lengths of strong-shear and strong-gradient neutral lines \citep{falcon2008}; spot areas, magnetic fluxes and average magnetic field \citep{wang2008};  the total unsigned magnetic flux and the total magnetic dissipation \citep{song2009}; number of singular points \citep{huang2010, yu2010a}.

Another group of the investigations focused on the role of the helicity in the magnetic field. \citet{labonte2007} determined a specific peak helicity flux as a necessary condition for the occurrence of an X-flare. \citet{wang2007} also took into account helicity injection among other dynamic conditions.

Besides the above features small scale structures are also considered as pre-flare conditions. Flare occurrence may also be expected from highly intermittent magnetic configurations, in other terms from turbulent structures, which can be described by fractal dimension analysis. \citet{abra2003} and  \citet{georg2005} reported characteristic variations shortly prior to flares, furthermore, \citet{crisc2009} found correlation between the generalized fractal dimension and the flare index. At the same time \citet{georg2012} pointed out that the fractality and turbulence properties do not distinguish flare-active and flare-quiet active regions. 

For the purpose of forecasting the structural analysis may be insufficient in itself, the fractality results show that the pre-flare dynamics may contain basic information about the probability of the flares. \citet{liu2008} introduced a quantity for the description of the magnetic field dynamics: $\bf{E} = \bf{u} \times \bf{B}$, where $\bf{u}$ is the velocity of the footpoints of the magnetic field lines and found that the footpoint motions may trigger flares. \citet{murr2012} reported further pre-flare changes: increase of the vertical field strength and current density as well as approaching of the magnetic flux towards the vertical by about $8^{\circ}$. 

All of the above mentioned methods are based on magnetograms. Our objective is a new type of measure for the description of nonpotentiality by using sunspot data. Sunspots are discrete entities instead of the continuous magnetic field distributions of the magnetograms but they are locations of high flux densities so they are presumably the dominant components of the processes. We analyse the pre-flare values and the behaviour of the horizontal magnetic field gradient in the area of the flare in order to find indicative values of this behaviour in the two-three days prior to the flare onset. The aim is to find signatures for the expected time of the onset of the flare and also for its intensity.

\section{Data and Analysis Tools}

\begin{figure}
\epsscale{0.8}
\plotone{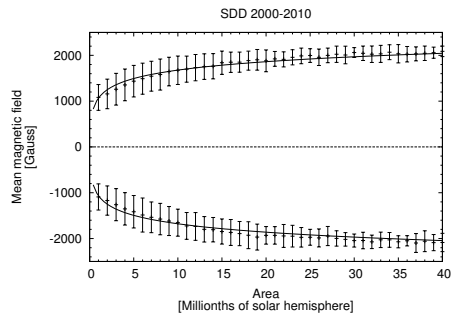}
\caption{Dependence of the mean magnetic field of both polarities in umbrae on the umbral area measured by SOHO/MDI on the solar disc center. The data are taken from the SDD catalogue.\label{calibr}}
\end{figure}

The most suitable dataset for this investigation is the SDD (SOHO/MDI-Debrecen Data), the most detailed sunspot catalogue covering the years of MDI operations, 1996-2010. It contains the data of positions, areas and mean magnetic fields for all observable sunspots and sunspot groups on a 1-1.5 hourly basis \citep{gyori2011}. This makes possible to follow the evolution of the internal magnetic configuration of the active regions in high spatial and temporal resolution. 
The SDD contains the mean field strengths of the sunspot umbrae and penumbrae. Figure~\ref{calibr} 
shows the dependence of the average of mean magnetic field on the umbral area within a distance of $10^{\circ}$ from the solar disc center. The sampling comprises 142,411 umbrae, the error bars represent the standard deviation of the mean magnetic field within the bins of umbral area of 1 MSH (millionth of solar hemisphere).

The curve fitted to the measured points can be described by the following formula:

\begin{equation}
B_{mean} \equiv f(A) = K_{1}\cdot ln(A) + K_{2}
\label{magnarea}
\end{equation}

\noindent 
where $|K_{1}|$ = 265 gauss and $|K_{2}|$ = 1067 gauss. This function renders a $\it B$ mean magnetic field to an $\it A$ umbral area. Considering that both the magnetic field and the area should be corrected for center-limb variation we decided to use directly the sunspot area data which are suitably corrected for geometrical foreshortening in the SDD. The targeted physical quantity is the horizontal gradient of the magnetic field between two areas of opposite magnetic polarities. The magnetic field will be represented by the total amount of magnetic flux enclosed within the umbra which can be obtained by multiplying the $B_{mean} \equiv f(A)$ mean magnetic field by the $\it A$ area. We define the following proxy measure to represent the magnetic field gradient between two spots of opposite polarities having $A_{1}$ and $A_{2}$ areas at a $\it d$ distance

\begin{equation}
G_{M} = \left | \frac {f(A_{1})\cdot A_{1} - f(A_{2})\cdot A_{2}}{d} \right |   
\label{proxy}
\end{equation}

The measured values are converted to SI units: Wb/m. We consider this quantity as a possible proxy of nonpotentiality at the photospheric level. It has the following advantages in comparison with the quantities defined directly on the magnetogram data. By tracking the pre-flare variations in real-time it is useful to have an as quick procedure as possible and the above  $G_{M}$ quantity only needs the determination of the area and distance data of some selected spots, this may need a couple of minutes with a moderate computing background. The other advantage is that the correction for center-limb variation is much more straightforward and reliable for geometric data than for magnetic data while the (1) calibration formula can (and probably should) be improved in the future.

\section{Pre-flare Dynamics of the Magnetic Field Gradient \\ Case Studies}

\begin{figure}
\epsscale{0.48} \plotone{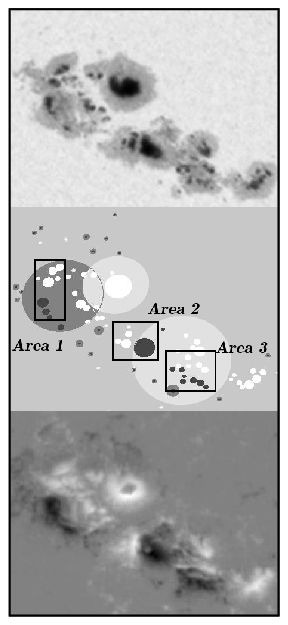}  \epsscale{0.7} \plotone{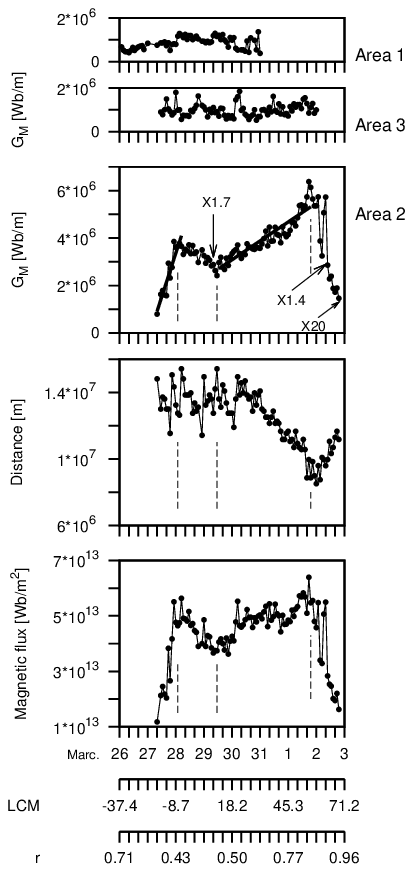}
\caption{Images: AR NOAA 9393 on 28 March 2001 at 06:23 UT: continuum image, reconstruction from SDD, magnetogram. Diagrams: variations of $G_{M}$ in areas 1 and 3;  diagrams of area 2: variations of $G_{M}$, distances, flux amounts. The two lower axes show the LCM and the radial coordinate (r) of area 2. \label{9393}}
\end{figure}

For case studies some energetic events have been selected to follow the development of the magnetic field gradient and to test the viability of the suggested approach. The images of Figure~\ref{9393} show the \objectname{NOAA 9393} active region on 29 March 2001, its white-light appearance, the magnetogram and (in the middle) the view of the sunspot group reconstructed from the SDD data. This panel is just a cartoon containing circles whose positions and areas correspond to the relevant data of spots in the catalogue and their grayscale color coding of the magnetic polarities corresponds to that of the magnetograms. Of course, the shapes of penumbrae cannot be recognized, they only represent the areas of penumbrae.

The active region produced numerous flares, only those cases are considered in the present work which are stronger than M5. We assumed that the most intense flares are in connection with the location of the strongest magnetic gradient and the variation in this region has been followed. This assumption is based on the finding of \citet{schrijver2007} that 'large flares, without exception, are associated with pronounced high-gradient polarity-separation lines'. The cartoon of the active region in Figure~\ref{9393} contains three selected areas containing spots of opposite polarities. The highest variability was exhibited by area 2. 

The upper two diagrams of Figure~\ref{9393} show the variations of  $G_{M}$ in the selected areas 1 and 3 between 25 March and 2 April. This can be regarded to be the behaviour of quiet areas. The third panel of the diagrams shows the variation of the $G_{M}$ flux gradient in area 2 between 27 March and 3 April in a cadence of 1.5 hours. In this area the spot pair with the highest $G_{M}$ is used. The times and intensities of the flares are indicated.

It is conspicuous that a steep rise and a high maximum value of the flux gradient is followed by a less steep decrease which ends with an energetic X1.7 flare on 29 March. Then a next gradient increase can be seen, a next high maximum value is followed by a steep decrease and two energetic flares with strengths X1.4 and X20.

Regarding to the (2) definition it is also informative to separate in this variation the roles of the amount of magnetic flux and the distance. The fourth diagram of Figure~\ref{9393} shows the variation of the distance between the spots involved in the highest $G_{M}$ value, the fifth diagram shows the variation of the amount of the involved magnetic flux. The corresponding time intervals are separated by dashed lines in the three lower diagrams. 

It can be seen that the starting rapid increase of the $G_{M}$ (first interval) and then the slower decrease (second interval) are due to the increase and decrease of flux amount, i.e. the strengthening and weakening of the spots. Then the longer strengthening (third interval) is due to increasing flux amounts and decreasing distances. After the highest maximum a medium intensity flare is released like in 28 March and the rapid decrease of $G_{M}$ (fourth interval) is resulted by decreasing flux amount and receding spots until the two X class flares.

The error of $G_{M}$ can be assessed as follows. Formula (1) contains three quantities, the $\it B$ magnetic field, (i.e. f(A)), the area $\it A$, and the distance $\it d$. The mean (typical) value of the magnetic field in the examined cases is 1600 Gauss, the mean error of the magnetic field rendered to a certain umbral area is $\Delta B_\mathrm{mean} \sim 250$ Gauss. This value is comparable to the errors presented by \citet{watson2011}, this has the largest contribution to the final error and it has the same constraint on all methods using MDI data. The mean umbral area is $3.8 \cdot 10^{13} \mathrm{m}^{2}$ its mean error is $3.8 \cdot 10^{12} \mathrm{m}^{2}$. The mean distance between the examined spot pairs is $ d \sim 2 \cdot 10^{7} \mathrm{m}$ its mean error is $\Delta d \sim 1.2 \cdot 10^{6} \mathrm{m}$ (considering that the error of position measurements is 0.1 heliographic degrees). By using mean values of these quantities and their mean errors and by applying the formula of the propagation of error one obtains a typical value for the error of inter-spot gradient: $\Delta G_{M} \sim 1.2 \cdot 10^{5} \mathrm{Wb/m} $. This value should be compared to the variations of $G_{M}$, see e.g. Figure~\ref{9393}: the mean fluctuation in the quiet areas 1 and 3 are about $ \sim 10^{6} \mathrm{Wb/m} $, in area 2 the decrease  is about $ \sim 1.5 \cdot 10^{6} \mathrm{Wb/m} $ and the two increases are more than $3 \cdot 10^{6} \mathrm{Wb/m} $. Thus the mean error is less than $10 \% $ of the addressed variations of $G_{M}$ which can be considered to be significant.

\begin{figure}
\epsscale{0.5} \plotone{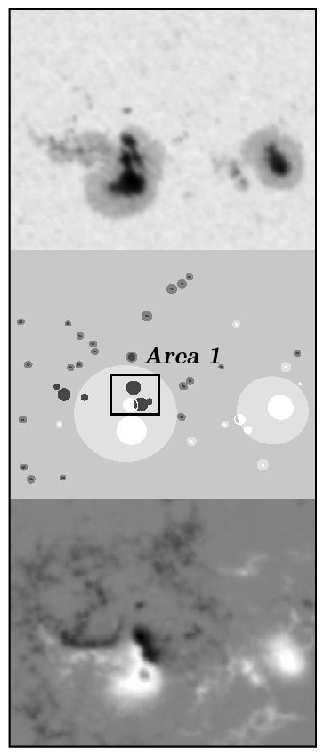}  \epsscale{0.7}  \plotone{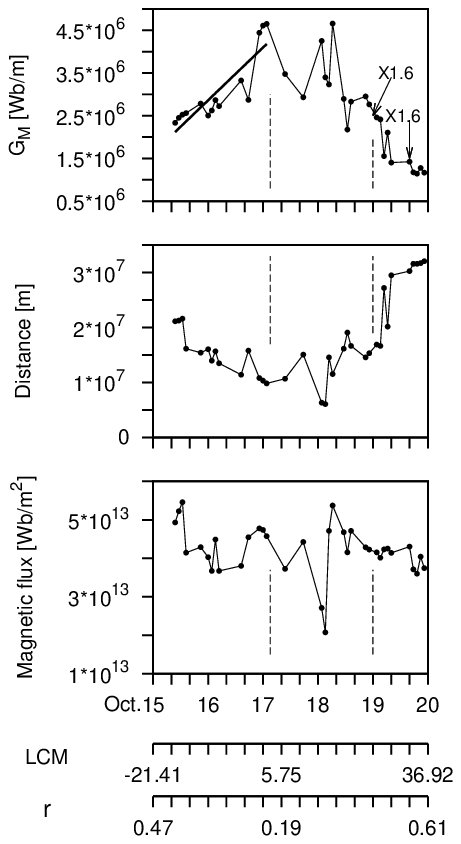}
\caption{The same images and diagrams for the active region NOAA 9661 as in Figure 2 for NOAA 9393. The images are taken on 17 Oct. 2001 at 16:37 UT.\label{9661}}
\end{figure}

The two similar variation patterns in area 2 prior to energetic flares may be signatures of the imminent eruption. This has been checked in another active region, the \objectname{NOAA 9661}, see Figure~\ref{9661}. The meanings of panels are the same as in Figure~\ref{9393} but without the comparison to quiet areas.

The events between 15-19 Oct. 2001 are as follows: the flux amount fluctuates but no continuous trend can be observed. The introducing increase of $G_{M}$ (first time interval until the first dashed line) is due to the approaching motion of the involved spots, then the decreasing trends until the two X1.6 flares (second and third intervals) are due to the receding motions of spots. It is conspicuous that in both presented cases the energetic flares are introduced by a certain decrease of the magnetic field gradient after a high maximum value.

\begin{figure}
\epsscale{0.5}  \plotone{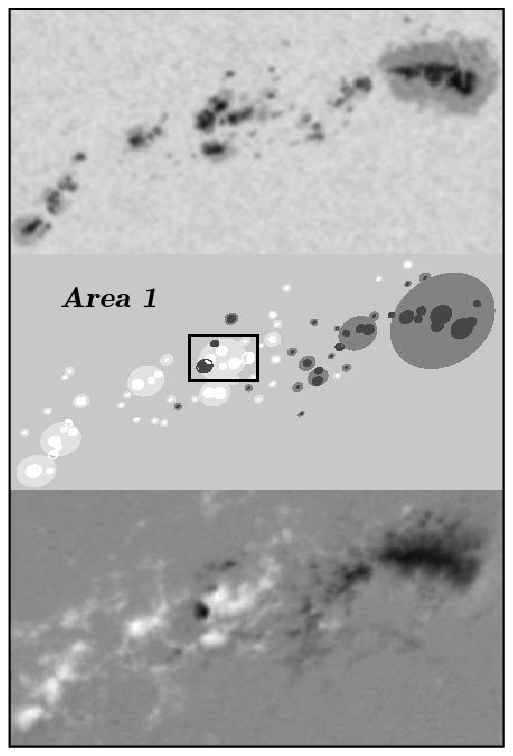}  \epsscale{0.7}  \plotone{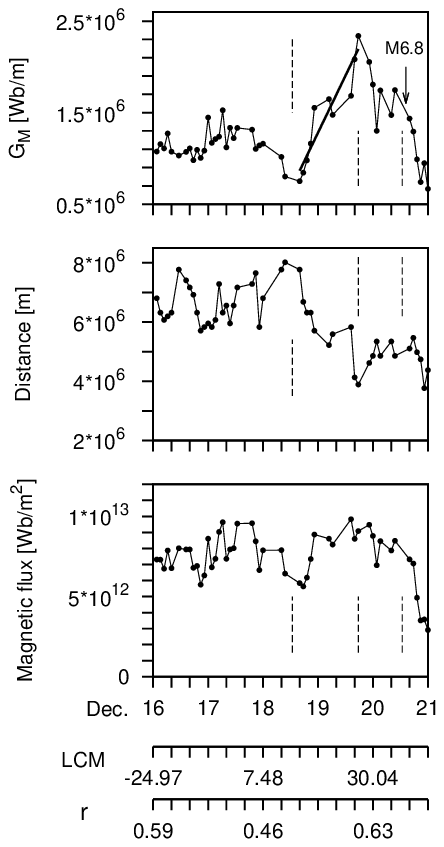}
\caption{Variations in the active region NOAA 10226 within the selected area 1 between 16 and 21 December, 2002, the meanings of the panels are the same as in Figures 2 and 3. The images show the active region on 19 December, at 16:54 UT.\label{10226}}

\end{figure}

The third example is the active region \objectname{NOAA 10226}, see Figure~\ref{10226}, the meanings of the panels are the same as in Figures~\ref{9393} and~\ref{9661}, the area of the highest flux gradient is indicated by a rectangle.

In this case the first gradient increase (until the second dashed line) is caused by both the approaching motions and strengthening of the spots. After the maximum the gradient decrease (after the second dashed line) is mainly due to the receding motions which end with an M6.8 flare.

\begin{figure}[!ht]
\epsscale{0.7}
\plotone{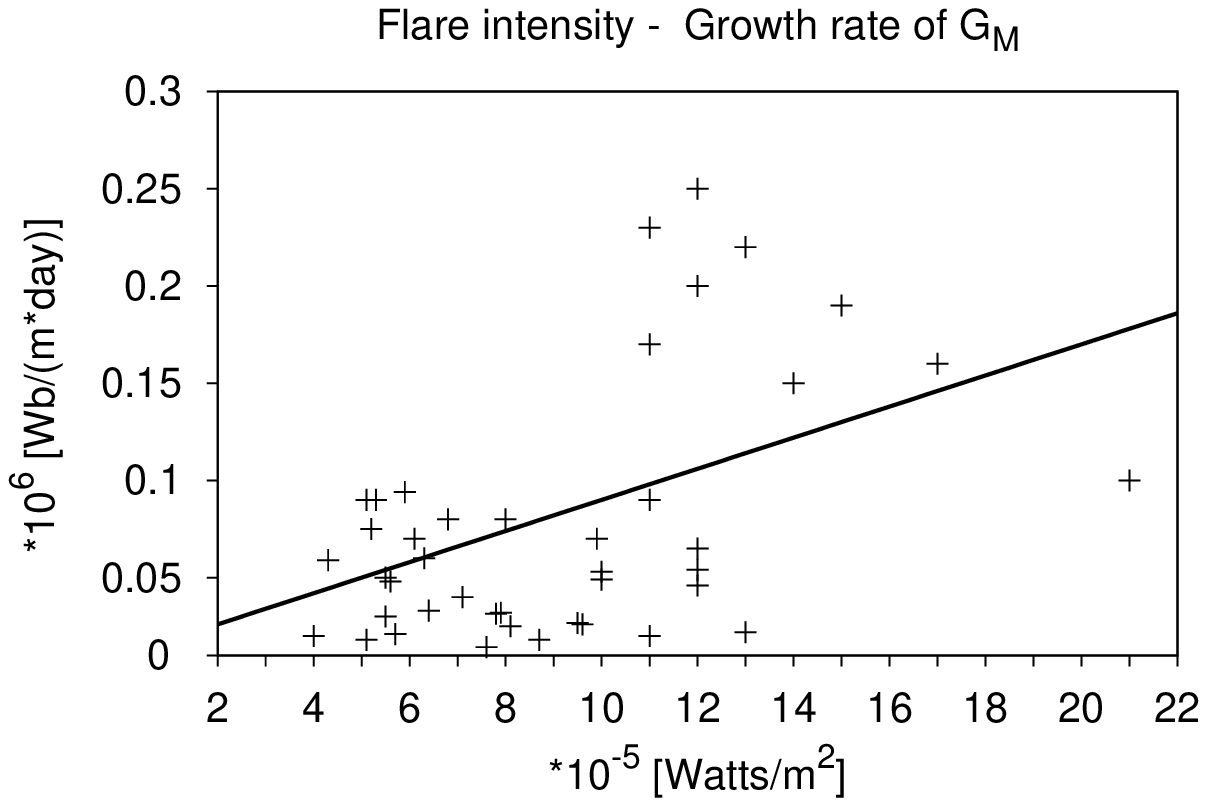} \\
\plotone{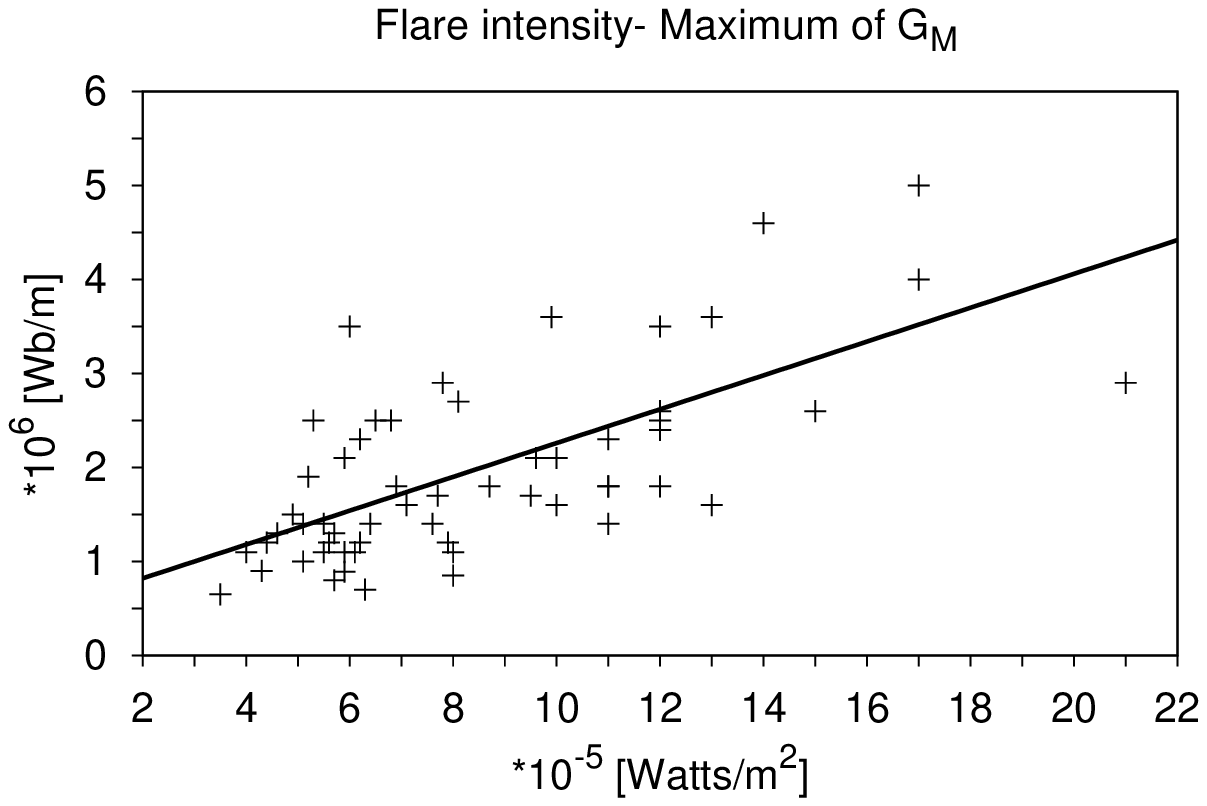}\\
\plotone{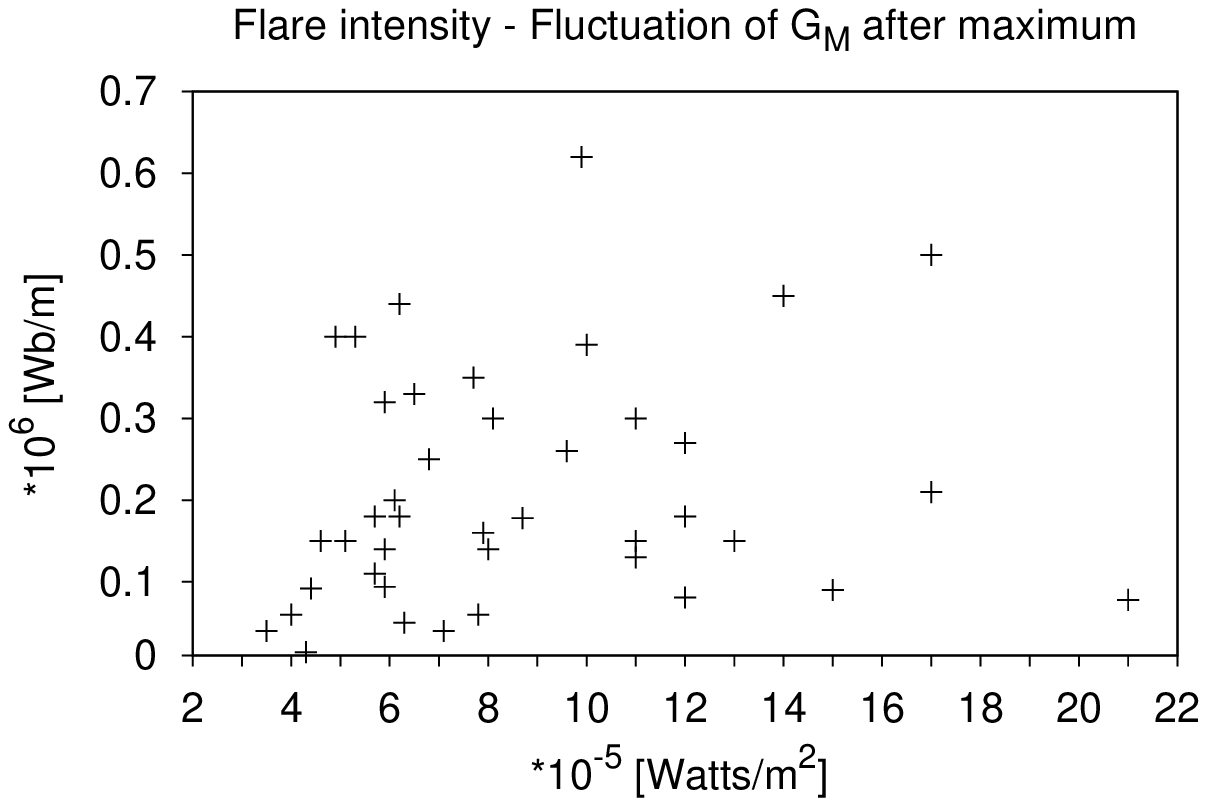} \\
\plotone{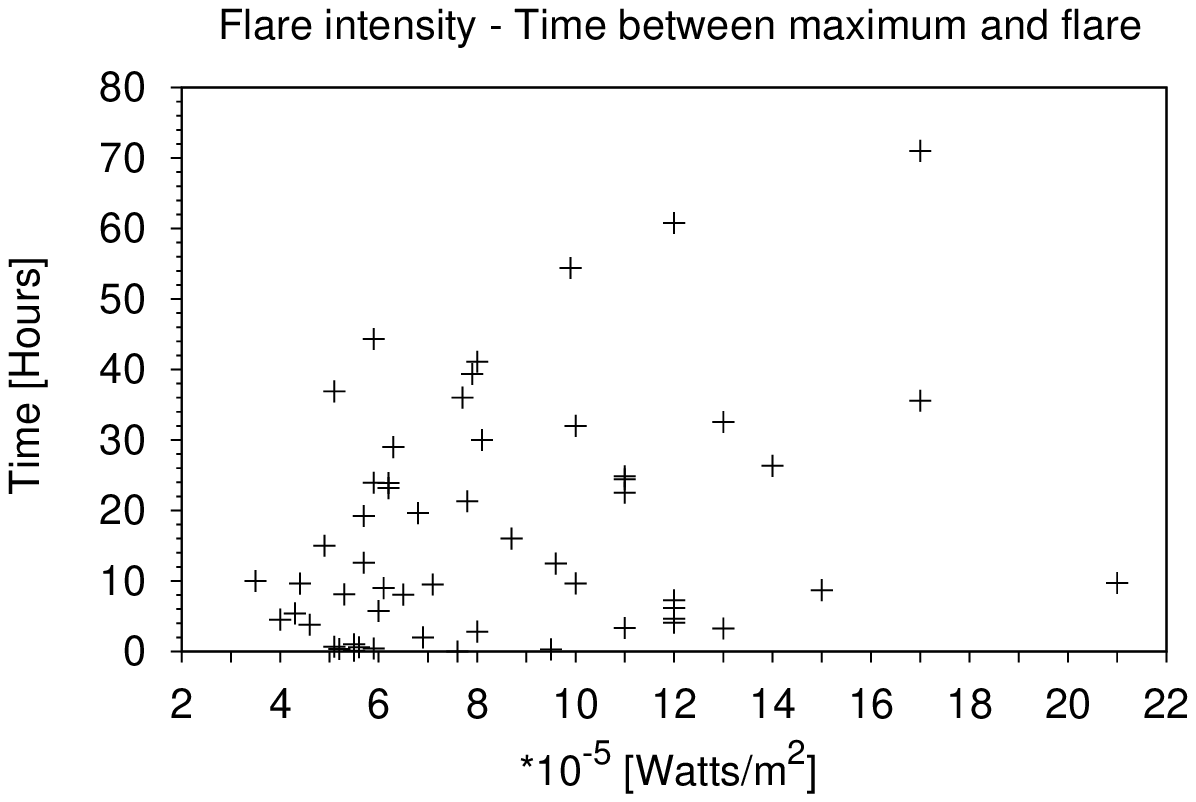} 
\caption{Relationships of the characteristics of flux gradient variations with the strengths of the produced flares: growth rate (upper panel), the achieved maximum (second panel), the fluctuation during its decrease (third panel) and the time between the maximum and flare onset (lower panel). The x-axes show the maxima of flare intensities in the 1-8 {\AA} wavelength range.\label{jel}}
\end{figure}

\section{Statistical Studies of the Variations}

In the above case studies the following properties of the pre-flare variations of the (2) magnetic flux gradient may be worth considering for further statistical investigations: i) at the beginning a steep rise ii) a high maximum, iii) decrease for several hours, iv) strong fluctuation during the decrease. All flares were introduced by this series of events.

The above characteristics have been examined on an appropriate sampling. The selection criteria were as follows. All groups produced a single significant flare after the maximum, i.e. either a single flare larger than M5 class, or an X-flare without preceding flares larger than M5. In a few cases the presented pre-flare behaviour of the $G_{M}$ was also recognizable before flares weaker than class M5 but the applicability of the method rapidly weakens toward less intense flares. The sample contains a few exceptions, five M4.x cases and a single M3.5 case, these were singular events in the given spot groups. The aim of these requirements is that the tracked variation of $G_{M}$ can unambiguously be rendered to a specific flare event. 

The flare should have been released no further than $-40^{\circ}$ to the East from the central meridian (in order to see its precursors). The entire targeted pre-flare variation should have taken place within $-70^{\circ}$ and $+70^{\circ}$ from the central meridian. Altogether 57 active regions fulfilled these requirements. The above restrictions mean that from the flares in Figures~\ref{9393}--~\ref{10226} only the X1.7 flare of AR NOAA 9393 is included in the statistics because it is only preceded by a moderate event, in the rest of the cases the energetic flares are either close to each other or beyond $+70^{\circ}$. However, the mentioned trends in the behaviour of $G_{M}$ can be recognized in these cases too.

Figure~\ref{jel} shows the relationships between the strengths of the produced flares and the following pre-flare characteristics:  the starting growth rate and the maximum value of the flux gradient, the time between the maximum and the flare after some decrease, and the strength of fluctuation during the decrease. This last quantity was computed as the standard deviation obtained by a linear fit to the data between the maximum and flare onset. Instead of the GOES-classification, the strengths of the flares are scaled in $\mathrm{watts/m}^{2}$.

\begin{figure}[!ht]
\epsscale{0.7}
\plotone{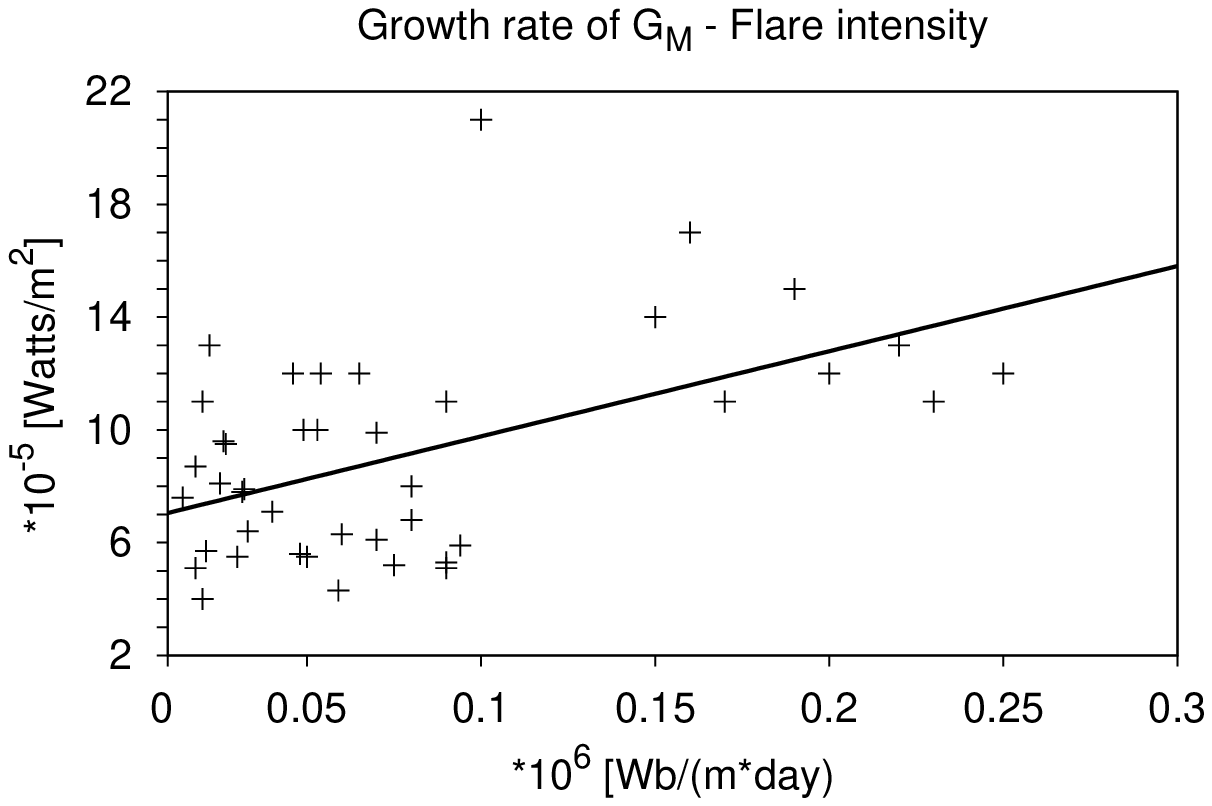} \\
\plotone{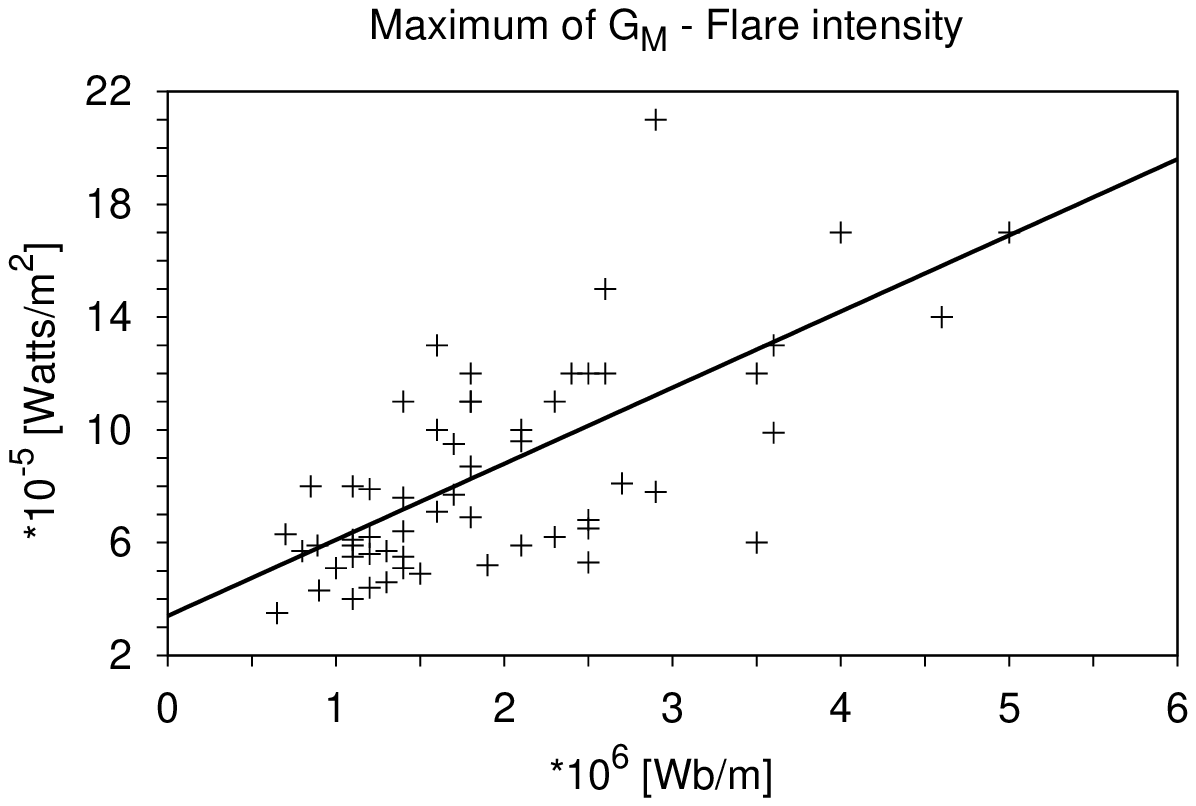}
\caption{Inverse plots of the two upper panels of Figure 6, the intensities of the produced flares in the 1-8 {\AA}  wavelength range with respect to the steepness of the rise of flux gradient until the maximum (left panel) and to the maximum value of the gradient (right panel).\label{inverse}}
\end{figure}

The relationship with the growth rate (Figure~\ref{jel} upper panel) is weak, a more unambiguous relationship can be seen in the second panel between the strength of the produced flare  and the maximum of the $G_{M}$ flux gradient. In order to get more suitable information for the causal connections between these quantities the inverse forms of these two relationships are plotted in Figure~\ref{inverse}.

The equation of the regression line in Figure~\ref{inverse} between the strength of the flare ($S_\mathrm{flare}$) and the maximum of the $G_{M}$ flux gradient ($G_{Mmax}$) can be written as

\begin{equation}
S_\mathrm{flare} = a\cdot G_{M\mathrm{max}} + b   
\label{flareint}
\end{equation}

\noindent
$ a = 2.7 \cdot 10^{-11} \pm 0.4 \cdot 10^{-11} (\mathrm{watts/m}^{2})/(\mathrm{Wb/m}) $ \\
$ b = 3.4 \cdot 10^{-5} \pm 0.7 \cdot 10^{-5}  \mathrm{watts/m}^{2} $

\begin{figure}
\epsscale{0.7}
\plotone{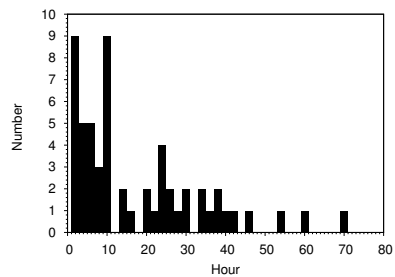}
\caption{Distribution of time delays between the achieved maximum of magnetic flux gradient and the flare onset in the 57 cases examined.\label{time}}
\end{figure}

The fluctuations and the delay times do not show any relationships with the strengths of the produced flares but the distribution of the delay times shows that in half of the cases the flares erupt within 10 hours after the maximum of $G_{M}$, see Figure~\ref{time}.

\section{Discussion}

The presented phenomena preceding the solar flares seem to raise new aspects of the pre-flare conditions. The study is based on the magnetic fields of sunspots, rather than distributions of magnetic fields in magnetograms. The spots are discrete entities of the active region magnetic fields, their high flux densities may render them the most efficient components of the pre-flare dynamics of active regions. The following features are of interest:

1) Quick growth of the $G_{M}$ magnetic flux gradient defined by (2). This is not surprising, this is the process of building up the nonpotential component of the active region magnetic field. The rise takes about 1.5-2 days, the steepness of the strengthening is only weakly indicative for the flare intensity.

2) High maximum of $G_{M}$ at about $3\cdot 10^{6}$ Wb/m. This is also obvious because this is the measure of nonpotentiality in the given region. This maximum value shows the only unambiguous relationship with the intensity of the released flare. Equation (3) gives a tool to estimate the intensity of the expected flare from the measured maximum of the flux gradient. This equation and the right panel of Figure~\ref{time} may be considered to be a relationship between the proxies of the free energy and the released energy.

3) Decrease of $G_{M}$ after maximum until the flare onset. This might be the most surprising pre-flare signature because one could expect that the state of strongest gradient is the most efficient trigger of the eruption. The after-maximum weakening prior to the flare may imply that the condition for the reconnection is not only the compression of the oppositely oriented flux ropes but also a subsequent loosening. 

4) The significant fluctuation during the decrease is also a characteristic feature of the pre-flare dynamics, but it is not related to the flare intensity. Its measured values are between $(3\cdot 10^{5}$ and $6\cdot 10^{5}$ Wb/m  during the decrease prior to the X-flares whereas it is  $7.5\cdot 10^{4}$ Wb/m in the quiet domain No.1 of active region \objectname{NOAA 9393} (Figure~\ref{9393}). 

5) The obtained characteristics are mostly signatures for major flares. The flare class M5 as a lower threshold may seem to be arbitrary, but it can be justified with Figure~\ref{jel}. Its upper right panel shows that at M4 class flares the maximum of $G_{M}$ is about $ 1\cdot 10^{6}$ Wb/m, while the lower left panel of Figure~\ref{jel} shows that the fluctuation of $G_{M}$ may be as high as $0.7\cdot 10^{6}$ Wb/m, thus the temporal variation of $G_{M}$ may be covered by the fluctuation in the case of weak flares. The recognizability of the presented beaviour decreases toward weaker flares at about $G_{M} \sim 1\cdot 10^{6}$ Wb/m and M4 class, it can be identified for flares more intense than M5.

6) In half of the examined cases the flare took place within 10 hours after the maximum of $G_{M}$.

The extension of the considered area to a central meridian distance of  $\pm 70^{\circ}$ may seem to go too far, \citet{schrijver2007} focuses on an area of $\pm 45^{\circ}$ by using LOS magnetograms, \citet{leka07} extend the region until $\pm 60^{\circ}$ by using vector magnetograms. In our case the applicability of the method until $\pm 70^{\circ}$ will only depend on the possibility to determine the polarity of the relevant spots, because the correction for the foreshortening of the umbral area can be made more reliably than for the center-limb variation of the magnetic field. The method is mostly limited at the eastern side because of a minimal necessary length of observable pre-flare evolution. 

The release of the magnetic free energy in current-carrying domains needs some external triggering or favourable conditions. The observed fluctuation of the $G_{M}$ seems to be an agent in reinforcing the reconnection event. What is more intriguing, however, the decrease after maximum can be an even more important process in making possible the reconnection. A possible interpretation considers the growing distance of the spots of opposite polarities. This may be the signature of new flux emergence which is an important element of pre-flare conditions according to several authors \citep{li2000a, li2000b, wang2007, lee2012, schrijver2005}, whereas this emergence injects helicity into the given domain of the magnetic field structure \citep{labonte2007, wang2007}. The injected helicity is also part of the nonpotential component of the active region magnetic field \citep{labonte2007}. \citet{luoni2011} pointed out that the specific elongations of the emerging bipolar regions, the "magnetic tongues", are observable signatures of the helicity carried by the emerging flux so the growing distance of the spots of opposite polarities may imply simultaneous flux emergence and helicity injection. 

Another possible explanation of this observed decrease is given by the theoretical study carried out by \citet{kusano2012}. They proposed that the reconnection on a current sheet is caused by the coronal diverging flows that remove magnetic flux and plasma from the reconnection site. If the photospheric diverging flow can work so, it could explain why the decrease of $G_{M}$ is associated with the trigger of flares. \citet{yamada1999} also mentions this process as a "pull mode" in laboratory experiments on reconnection. 

Equation (3) shows that the gauge of nonpotentiality makes possible to estimate the expected flare energy. \citet{falcon2009} found that the free energy content of any active region has a maximum attainable level which is in relationship with the total magnetic flux, the plot of this relationship is called "main sequence" by the authors. The produced flares and CMEs are fed from this free energy reservoir. If the active region is close to the free energy limit then flare eruption and CME are highly probable. It should be admitted that the presented statistical results refer to a single flare in an active region or the first member of a series of flares. Falconer et al. (2012) showed that previous major flares may modify the forecast chance.

The investigations listed in the introduction use different properties of the active region magnetic fields to assess the probability of flare onset \citep{schrijver2009} and the validity of their predicting capabilities may also be different. The spatial range of the considered object seems to be related to the temporal range of the prediction capability. The longest timescale belongs to the cyclic variation, the shortest timescale, 15-20 minutes is typical in the studies using the quick variations in the fine structure of the magnetic field, the fractality \citep{abra2003}. The analyses of the entire active regions on the magnetograms focusing on the neutral line or the gamma-configuration are capable to foresee and estimate the flare probability for a couple of days ahead.

\acknowledgments

The authors are indebted to Kanya Kusano for kind encouragements and helpful comments. The research leading to these results has received funding from the European Community's Seventh Framework Programme (FP7/2012-2015) under grant agreement No. 284461 (eHEROES project).

\end{document}